\documentclass[twocolumn]{aastex63}

\usepackage{amsmath}

\shorttitle{Fundamentalization of classical Cepheid periods}
\shortauthors{Pilecki B.}
\graphicspath{{figs/}{./}}

\begin{document}

\title{Fundamentalization of periods for first and second-overtone classical Cepheids
}

\correspondingauthor{Bogumił Pilecki}
\email{pilecki@camk.edu.pl}

\author[0000-0003-3861-8124]{Bogumi{\l} Pilecki}
\affiliation{Centrum Astronomiczne im. Miko{\l}aja Kopernika, PAN, Bartycka 18, 00-716 Warsaw, Poland}


\begin{abstract}

Almost half of all classical Cepheids do not pulsate in fundamental mode, and nowadays, the fundamentalization of their higher-mode periods is frequently applied to increase the sample size in astrophysical investigations and allow for comparison with fundamental-mode Cepheids. On the other hand, the relations used to obtain fundamentalized periods are either old or based on small samples that cover narrow period ranges.
We used available data of 989 Cepheids pulsating in at least two modes to obtain modern, high-quality empirical fundamentalization relations applicable in a wide range of periods of first- and second-overtone Cepheids for metallicities typical for the Milky Way and Magellanic Clouds. A clear correlation between the features of these relations and metallicity is seen, and periods with lower sensitivity to metallicity are identified. We also compare our results with double-mode Cepheids from the M31 and M33 galaxies. For the first galaxy, this indicates Cepheids have metallicities from supersolar to typical for the LMC, while for the latter, from solar to typical for the SMC. A general discussion of the usage of a different type of fundamentalization relations depending on the scientific problem is included.



\end{abstract}



\section{Introduction} \label{sec:intro}

Classical Cepheids (hereafter also Cepheids) are crucial for various fields of astronomy, including stellar oscillations and the evolution of intermediate and massive stars, and have an enormous influence on modern cosmology \citep[see, e.g., the review of][]{Bono_2024_Review_Cepheids}.  Since the discovery of the relationship between their pulsation period and luminosity (the Leavitt Law, \citealt{1912HarCi.173....1L}), the Cepheids have been extensively used to measure distances in the Universe \citep{Anderson_2024book_Cepheid_dist_H0}. They are radially pulsating evolved intermediate and high-mass giants and supergiants, mostly located in a well-defined position on the helium-burning loop (called the {\em blue loop}). About 43\% of Cepheids do not pulsate in fundamental (F) mode, being mostly either first-overtone (1O) or second-overtone (2O) pulsators \citep{Soszynski_2017AcA_OCVS_MC_Cep, Udalski_2018_GalCep_EclBin}. To increase the significance of analysis, it is therefore often necessary to combine the samples and, for that, to fundamentalize the higher-order mode pulsation periods \citep[e.g.,][]{Moskalik_2005_angdiam_cep, Evans_2013_orbperiods, Marconi_2017_SMC_Cep_models, cepheid_astrophysics_2018, Breuval_2020_MW_PLRs_GDR2, Bhardwaj_2024_cepheid_metal_PL}. Period fundamentalization is also used to apply characteristics of fundamental-mode Cepheids to the first-overtone ones \citep{Genovali_2014_MWthin_cep_metal_grad, Evans_2015AJ_cep_binarity_30perc} or to compare the Cepheids pulsating in different modes \citep{Gallenne_2018_V1334Cyg_Cep}.

There are different empirical approaches for the fundamentalization of pulsation periods. A classical method uses double-mode Cepheids (also called beat Cepheids), which pulsate in both the F and 1O mode, to find a relation between the period ratio $P_{1O}/P_{F}$ and other parameters, in principle one of the periods and metallicity \citep{Alcock_1995_double_cepheids, Feast_1997_fundper, Kovtyukh_2016_fundper_feh, Sziladi_2018_feh_per_prat}. Such a method should be consistent with the period-mass-radius (PMR) relation \citep{Bono_2001APJ_cep_MR_mass_discrep, cepheid_astrophysics_2018} but not with the period-luminosity one, being mostly insensitive to the temperature change across the instability strip (IS). Consequently, when such fundamentalized periods are used, 1O Cepheids lie slightly above the period-luminosity (P-L) relation for F-mode Cepheids. The difference increases as we move from the central part of the instability strip occupied by double mode F+1O Cepheids, the so-called OR region \citep[see, e.g.,][]{Bono_2001_2O_Ceps_SMC, Buchler_2009_AIPC}, towards the blue edge of the IS. 
An alternative method that, on average, keeps the luminosity (but does not follow the PMR relation) was developed by \citet[hereafter: P21]{cepgiant1_2021} to look for overbright Cepheids regardless of their mode.
This was obtained by minimization of scatter in a joint (F+1O) P-L relation, where 1O periods were fundamentalized using a fitted function. This method corrects for the average difference in luminosity between the F and 1O Cepheids along the line of the same period in the IS. Therefore, the luminosity obtained for 1O Cepheids from P-L relations for F-mode Cepheids (using the fundamentalized period) should be, on average, the same as the measured one.
And finally, period ratios that are predicted from theoretical modeling for a given metallicity \citep[e.g.,][]{Bono_2001_2O_Ceps_SMC} can also be used for fundamentalization of higher-order mode periods.

This paper presents modern, high-quality empirical fundamentalization relations based on double-mode Cepheids that are applicable in a wide range of periods for metallicities ([Fe/H]) typical for the Milky Way (hereafter also MW), Large Magellanic Cloud (LMC), and Small Magellanic Cloud (SMC).

\section{First-overtone mode}
\label{sec:first}

About 42\% of Cepheids pulsate in at least 1O mode, not having the F mode excited. In this section, we derive equations for the fundamentalization of 1O periods. 

\subsection{Data}
 From the OGLE-4 catalogs for the Milky Way \citep{Udalski_2018_GalCep_EclBin} and Magellanic Clouds \citep{Soszynski_2017AcA_OCVS_MC_Cep} we retrieved pulsation periods for 231 Cepheids pulsating at least in fundamental and first-overtone modes and calculated for them the corresponding period ratios, $P_F/P_{1O}$. The MW sample was extended with 18 Cepheids listed in \citet{Sziladi_2018_feh_per_prat}. In total, in the analysis, we used 101 objects from the LMC, 69 from the SMC, and 79 from the MW. For all occurrences, the unit for periods is days.

\begin{figure}
  \centering
  \includegraphics[width=0.98\linewidth]{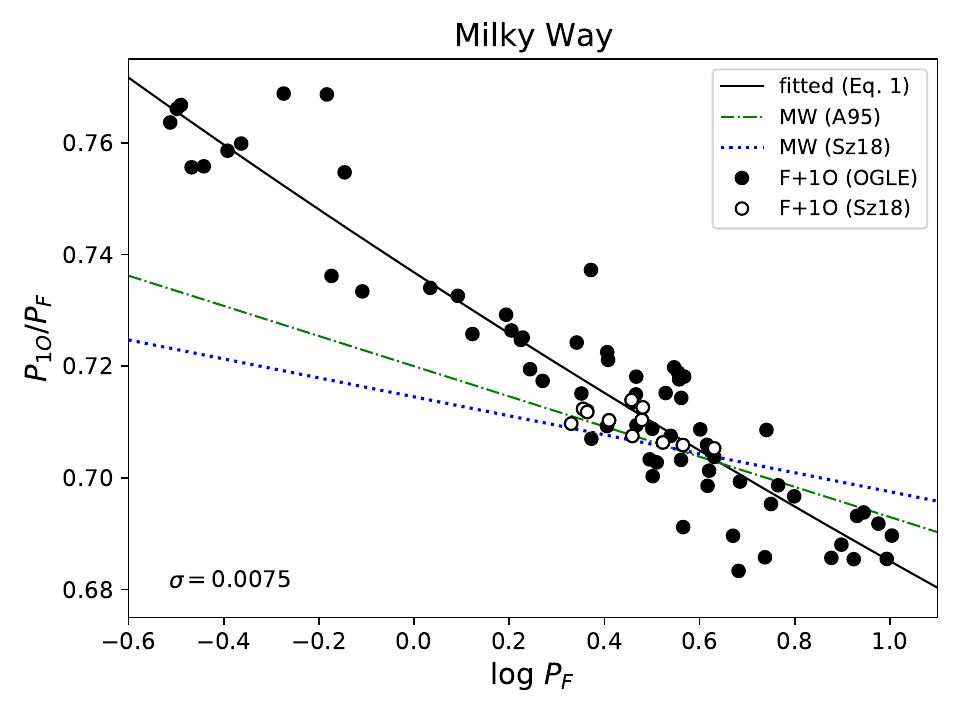}
  \includegraphics[width=0.98\linewidth]{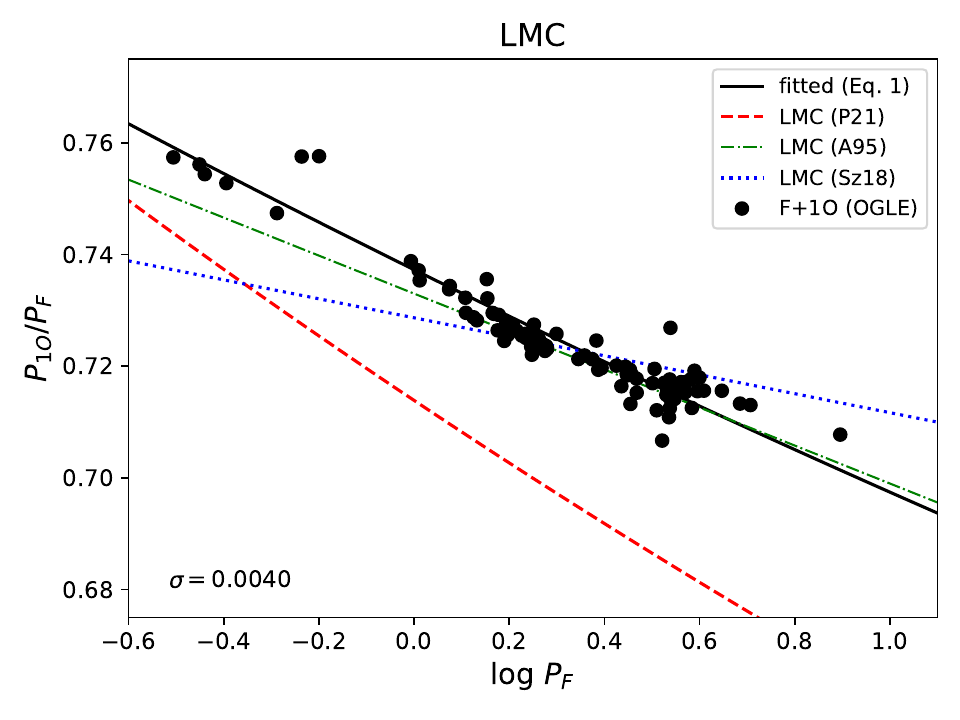}
  \includegraphics[width=0.98\linewidth]{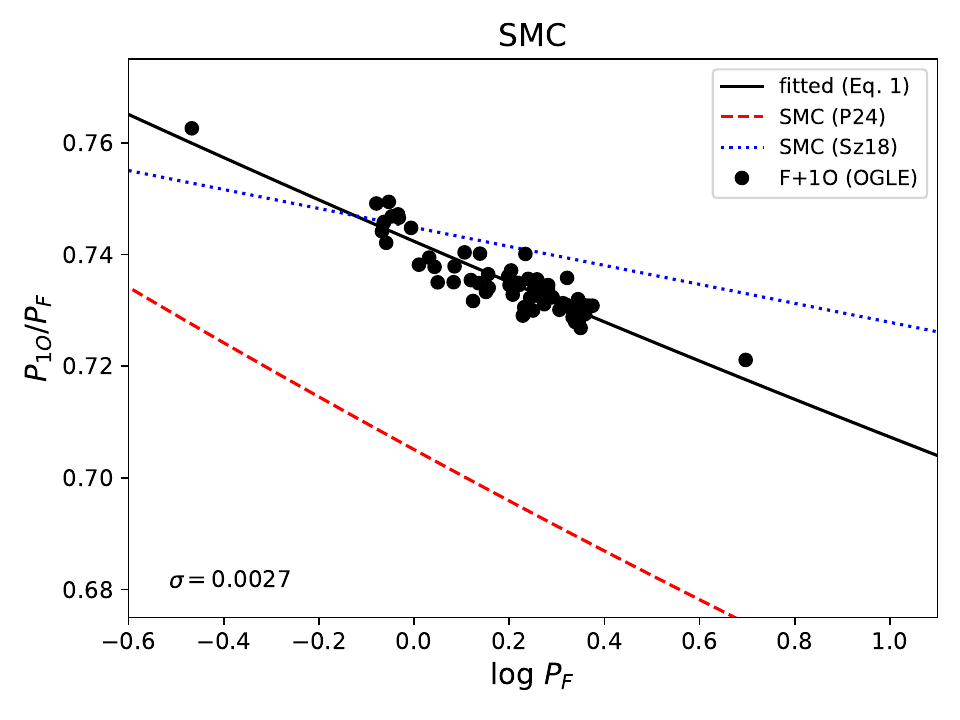}
  \caption{Period ratios for double-mode (F+1O) Cepheids from the OGLE-4 catalog for three considered galaxies. Black lines are the best linear fit for the data. Green dashed lines represent relations from A95. Blue dotted relations were taken from Sz18 for adopted average [Fe/H] of 0.0, -0.35, and -0.75~dex for MW, LMC, and SMC, respectively.
  Red dashed relations (P24) are not based on double-mode Cepheids (see text) but are shown here for comparison. The ranges of the X and Y axes are the same in all panels.}
  \label{fig:fund1}
\end{figure}

\subsection{Relations}
\label{sec:F1O_relations}
 For Cepheids of each galaxy, we fitted a relation in the form of $P_F/P_{1O} = a + b \log P_{1O}$, the same as in equation~1 of \citet[hereafter: P24]{Pilecki_2024_cepgiants2}, that makes the fundamentalization much easier than a typically provided $P_{1O}/P_{F} = a + b \log P_{F}$.  However, the used data and the best fit for the Milky Way and Magellanic Clouds are shown in Fig.\ref{fig:fund1} in a standard Petersen diagram $P_{1O}/P_{F}$ vs. $P_F$. 
 For comparison, we also show similar relations from \citet[hereafter: A95]{Alcock_1995_double_cepheids}, metallicity-dependent relations from \citet[hereafter: Sz18]{Sziladi_2018_feh_per_prat}, and those from P24. We note that the latter were obtained differently, using the Wesenheit indices of single-mode 1O and F-mode Cepheids as constraints. The best-fitting relations for double-mode Cepheids for each galaxy are given below:

\begin{subequations}
\begin{eqnarray}
\label{aeq:fund}
\ (MW)\quad P_F/P_{1O} =  1.371 + 0.106 \log P_{1O},\\
 (LMC)\quad P_F/P_{1O} =  1.367 + 0.079 \log P_{1O},\\
 (SMC)\quad P_F/P_{1O} =  1.356 + 0.068 \log P_{1O}.
\end{eqnarray}
\end{subequations}

These relations can be directly used to fundamentalize periods of 1O Cepheids. The high scatter of period ratios for the Milky Way is probably due to the metallicity spread \citep{Genovali_2014_MWthin_cep_metal_grad, Luck_2018_MW_cep_metal_multiphase_grads, Trentin_2024_c_metall_VI_grads}, and a more complex relation would be preferred.
However, the metallicity-dependent relation from Sz18 has a slope significantly different from the one measured here. The reason for this is probably a much lower period range of Cepheids with measured metallicity, which does not provide a good constraint for the $b \log P_{1O}$ term.
As a result, this relation does not reproduce correct period ratios for the lowest and highest pulsation periods for any of the considered galaxies. We note that the assumed metallicities do not affect this conclusion; they can only shift the relation vertically but not change the slope. We used metallicities [Fe/H] of 0.0 dex, -0.35 dex, and -0.75 dex for MW, LMC, and SMC, respectively, consistent with the homogenous metallicity study of Cepheids by \citep{Romaniello_2008_MW_LMC_SMC_cep_metal} and values adopted by \citet{Gieren_2018_PL_FeH} that considered several different estimates.
The LMC relation from A95, based on a larger number of Cepheids, is mostly consistent with our result, but a slight difference is notable at the shortest periods. Data from a wide range of periods are thus crucial to obtaining relations that are universally applicable, regardless of  Cepheid properties.
As expected, the relations from P24 give longer fundamentalized periods, compensating for the higher temperature and luminosity of 1O Cepheids. The corresponding magnitudes from the two approaches differ by about 0.03 to 0.07 mag, depending on the period.

\begin{figure}
    \centering
    \includegraphics[width=\linewidth]{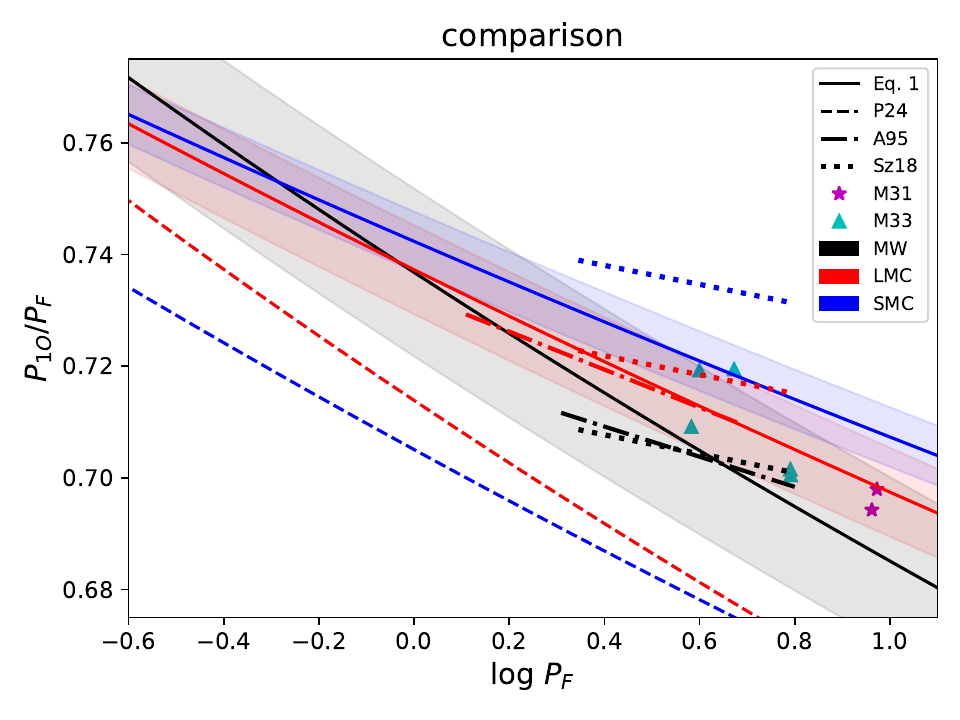}
    \caption{Comparison of the fitted relations for all considered galaxies. Shaded areas mark the 2-$\sigma$ range around them. The relations of A95 and Sz18 are plotted here only in their validity range, $\log P_0 =$ 0.1 -- 0.7 for the A95 LMC relation and 0.3 -- 0.80 for the rest. The position of known beat Cepheids in the M31 and M33 galaxies is also shown.}
    \label{fig:fundcomp}
\end{figure}

In Fig.~\ref{fig:fundcomp}, we compare the relations for all three galaxies.  The difference between them most probably comes from a different average metallicity of MW, LMC, and SMC, which is reflected by its correlation with the slope. Interestingly, at the short-period end, these relations either cross (MW with LMC and SMC) or get very close to each other (LMC with SMC).  Apparently, the dependence of $P_F/P_{1O}$ on metallicity is lower for periods shorter than one day, although it is not negligible (the scatter for MW Cepheids is significant there).
For longer periods, the difference increases considerably. As position on this diagram may be used to estimate the metallicity of Cepheids, we show here also the position of known beat Cepheids from M31 \citep{Poleski_2013_M31_beat_cepheids} and M33 \citep{Beaulieu_2006_M33_beat_cepheids} galaxies. For Cepheids in M31, this indicates the metallicity typical for the LMC, and for M33 Cepheids, a spread from solar to typical for the SMC.

\section{Second-overtone mode}
\label{sec:second}
About 7\% Cepheids pulsate in at least 2O mode without having the F mode excited. In this section, we derive equations for transforming 2O periods to their equivalents in 1O and F modes.

\subsection{Data}
We selected 740 Cepheids pulsating in at least the first and second overtone modes from the same OGLE-4 catalogs \citep{Soszynski_2017AcA_OCVS_MC_Cep, Udalski_2018_GalCep_EclBin} as used in the previous section. We then calculated the corresponding period ratios, $P_{1O}/P_{2O}$.
In total, in this part of the analysis, we used 329 objects from the LMC, 240 from the SMC, and 171 from the MW.

\begin{figure}
    \centering
    \includegraphics[width=\linewidth]{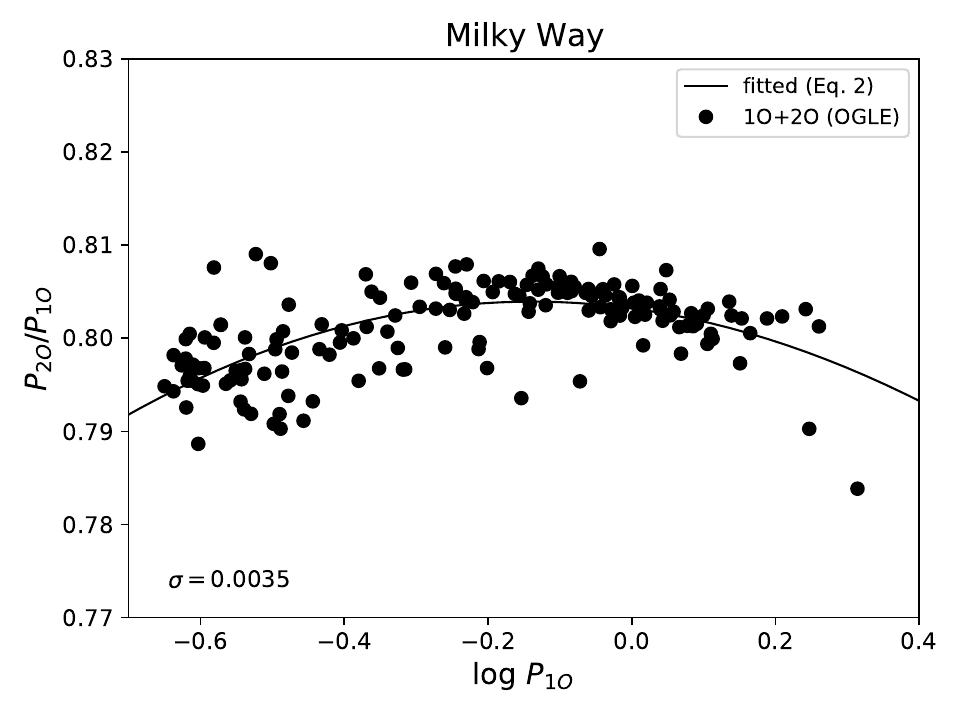}
    \includegraphics[width=\linewidth]{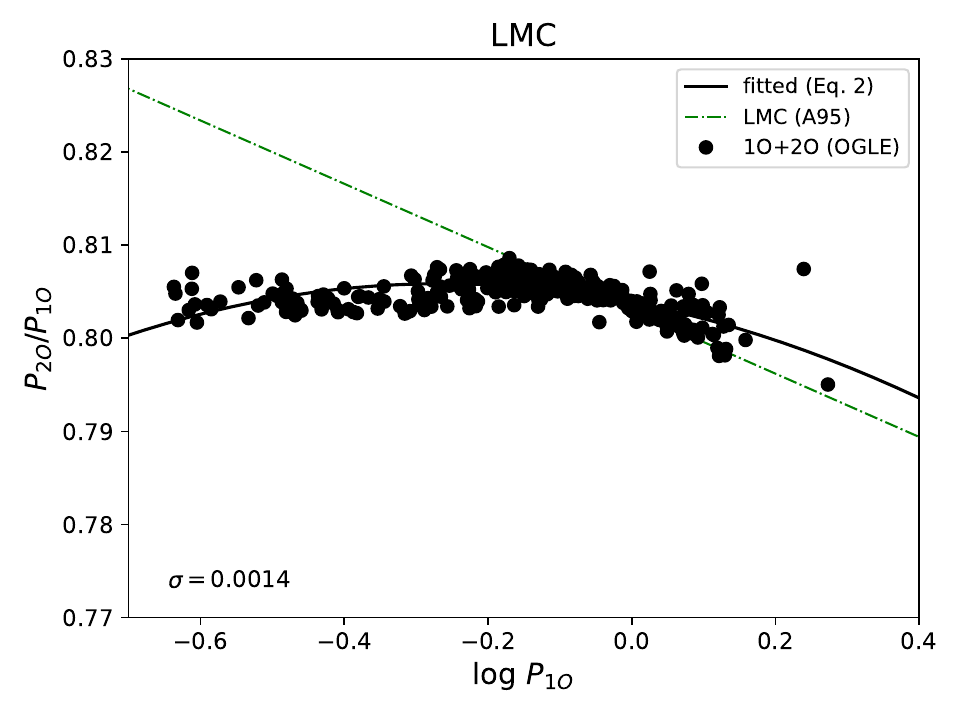}
    \includegraphics[width=\linewidth]{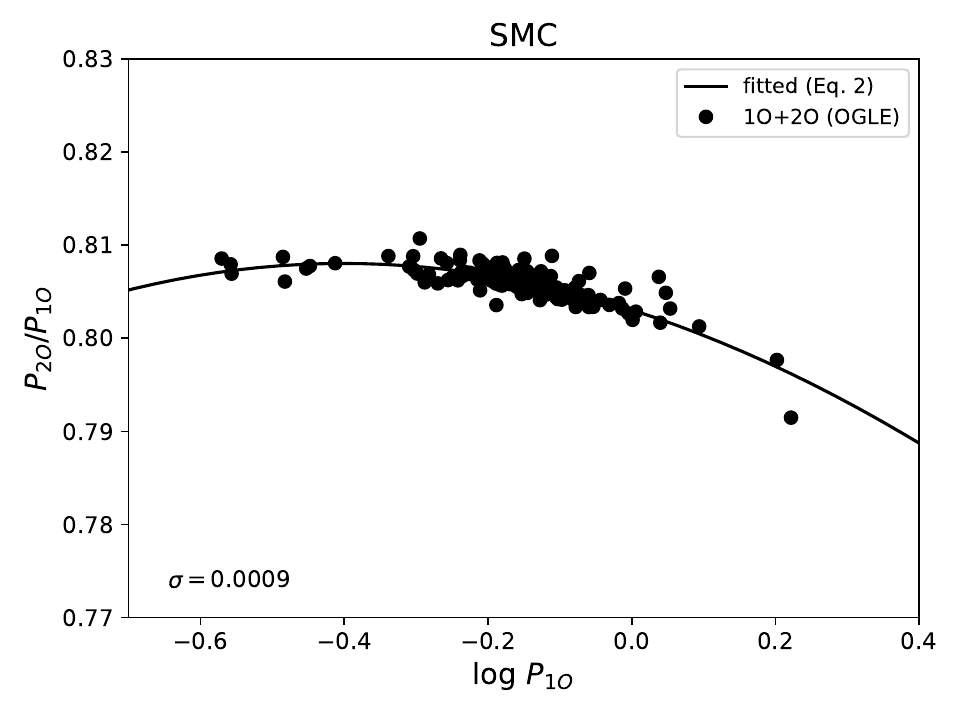}
    \caption{Period ratios for double-mode (1O+2O) Cepheids from the OGLE-4 catalog for MW, LMC and SMC. Black lines are the best parabolic fit for the data. The green dashed line represents the relation from A95. For MW, the data are generally highly scattered except around $P_{1O}$ = 1 day. The ranges of the X and Y axes are the same in all panels.}
    \label{fig:fund2}
\end{figure}

\subsection{Relations}

For Cepheids of each galaxy, we fitted a relation in the form of $P_{1O}/P_{2O} =  a + b \log P_{2O} + c \log^2 P_{2O}$.  These data are shown in Fig.~\ref{fig:fund2} together with the best fit for the Milky Way and Magellanic Clouds. For comparison, we also show a similar relation from A95.
The best-fitting relations for the MW (2a), LMC (2b), and SMC (2c) galaxies are given below:

\begin{subequations}
\begin{eqnarray}
\label{eq:second}
\frac{P_{1O}}{P_{2O}} =  1.247 + 0.028 \log P_{2O} + 0.059 (\log P_{2O})^2\\
\frac{P_{1O}}{P_{2O}} =  1.247 + 0.032 \log P_{2O} + 0.044 (\log P_{2O})^2\\
\frac{P_{1O}}{P_{2O}} =  1.249 + 0.048 \log P_{2O} + 0.048 (\log P_{2O})^2
\end{eqnarray}
\end{subequations}

These relations can be used to convert 2O periods to their 1O equivalents. Similarly to the  $P_F/P_{1O}$, a high scatter can be seen for the Milky Way. The A95 relation is clearly discrepant for low periods but fits the data reasonably well for the period range used to obtain it, i.e., $\log P_0 =$ -0.2 -- 0.1. As in the previous section, using data with a wide period range was important to obtain relations applicable regardless of the Cepheid period.

Although parabolic functions fit the data well, it seems that a linear function with a break would fit the data better for the LMC. However, we prefer to have the same formula for all relations and keep the number of fitted parameters to a minimum.
The analysis of these relations shows that the position of the maximum of $P_{2O}/P_{1O}$ shifts to shorter periods for decreasing metallicity, being located at $\log P_M =$ -0.139, -0.263, and -0.401 for the MW, LMC, and SMC, respectively. Using [Fe/H] values adopted in Section~\ref{sec:F1O_relations}, this gives a linear relation: [Fe/H] $= 0.40 + 2.9 \log P_{M}$.

\begin{figure}
    \centering
    \includegraphics[width=\linewidth]{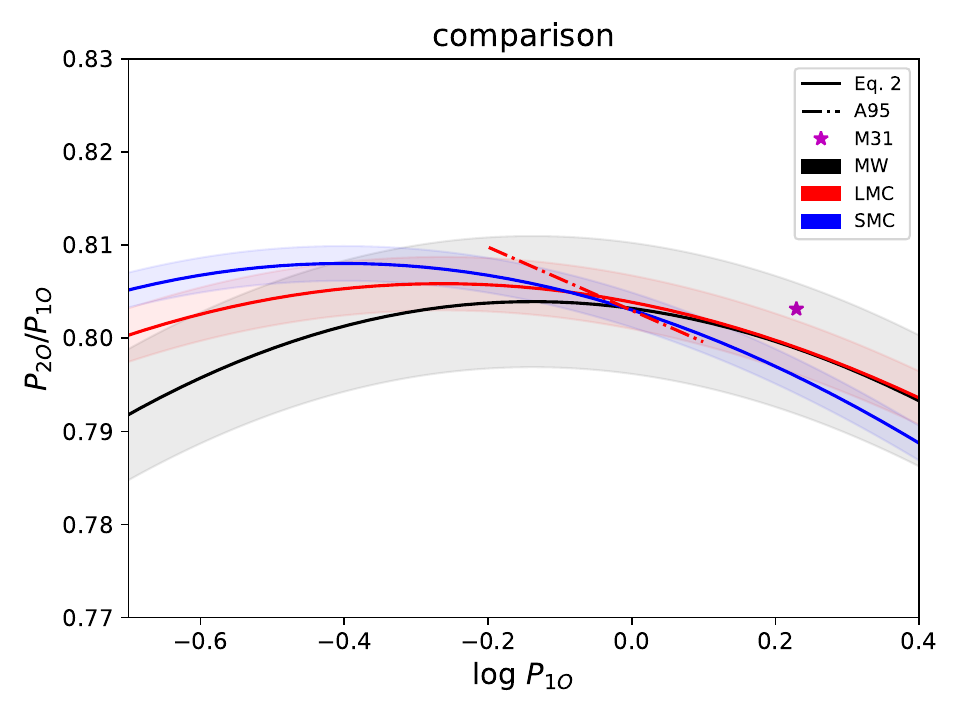}
    \caption{Similar to Fig.~\ref{fig:fundcomp} but for Cepheids pulsating in 1O and 2O modes. The fitted relations cross at around $P_{1O} =$ 0.9 days. The relation of A95 is plotted here only in its validity range, $\log P_0 =$ -0.2 -- 0.1. The position of one M31 beat (1O+2O) Cepheid is shown. }
    \label{fig:firstcomp}
\end{figure}

In Fig.~\ref{fig:firstcomp}, we compare the relations for all three galaxies. They cross at moderate periods and are similar for longer periods, while the discrepancy increases significantly at the shorter-period end.
From this comparison, we can infer that for $P_{1O}$ between 0.7d and 1.4d, $P_{1O}/P_{2O}$ have low dependence on metallicity. Actually, at the period of 0.9 days, these ratios are almost completely insensitive to [Fe/H], which is also reflected in a very low scatter for MW Cepheids around this value (see Fig.~\ref{fig:fund2}). We also overplot in this diagram the position of a known beat M31 Cepheid  \citep{Poleski_2013_M31_beat_cepheids}. A comparison with the presented relations indicates that it may have solar or supersolar metallicity.

\subsection{Fundamentalization}

A combination of relations given by Eq.~1 and 2 can be used to obtain fundamentalized periods for 2O Cepheids.
Below, we provide such a transformation in the same form and order as Eq.~2.

\begin{subequations}
\begin{eqnarray}
\label{eq:f2O}
\frac{P_{F}}{P_{2O}} =  1.723 + 0.171 \log P_{2O} + 0.081 (\log P_{2O})^2\\
\frac{P_{F}}{P_{2O}} =  1.713 + 0.144 \log P_{2O} + 0.062 (\log P_{2O})^2\\
\frac{P_{F}}{P_{2O}} =  1.702 + 0.152 \log P_{2O} + 0.068 (\log P_{2O})^2
\end{eqnarray}
\end{subequations}

To test these relations, we calculated fundamental-mode periods ($P_F^{rel}$) using $P_{2O}$ for the only four Cepheids in our sample that pulsate simultaneously in F and 2O modes (all of them are triple-mode F/1O/2O Cepheids). For OGLE-GD-CEP-1011 and OGLE-GD-CEP-1704 , we obtained a relative difference $\Delta_{rel}=(P_F-P_F^{rel})/P_F$ of 0.71\% and -0.24\%, respectively. For OGLE-LMC-CEP-1378 and OGLE-LMC-CEP-4718, the differences are 0.54\% and 0.03\%. All four values are below 1$\sigma$ for $\Delta_{rel}$ obtained for the corresponding relations for 1O Cepheids shown in Fig.~\ref{fig:fund1}.

Please note, however, that 2O Cepheids barely overlap in the IS with those pulsating in the F mode. Therefore, although the relations are well-defined in corresponding period ranges, such a fundamentalization of 2O periods may be considered a significant extrapolation.

\section{Final remarks}

Double-mode Cepheids are a subset of all Cepheids; they occupy a limited part of the instability strip, called the OR regions \citep{Bono_2001_2O_Ceps_SMC}. The presented relations are thus only an approximation that may worsen the farther we move from the source data in the parameter space. For example, they do not take the temperature dependence into account.

Moreover, we want to highlight that which empirical approach for fundamentalization should be applied depends on the objective for which the resulting data will be used. For example, the fundamentalization presented here should be preferred whenever we are interested in obtaining the expected physical properties of the stars. On the other hand, the transformations based on P-L relations, as in P21 and P24, should be used if we want to unify the samples of higher-order-mode Cepheids with the fundamental-mode ones or, in general, when we want to ignore the average difference in their temperatures and luminosities. An example of using both approaches for the same sample depending on the objective can be found in P24.

Alternatively, theoretical hydrodynamical models that account for a time-dependent treatment of convective transport \citep{Bono_2000_ApJ_CEP_puls_model_III, Marconi_2005_cepheids_puls_mod_met, sm08} offer the calculation of periods of different modes even for single-mode Cepheids, making it possible to obtain the fundamental mode period directly. This is equivalent to the first approach with the advantage of avoiding extrapolation. On the downside, to properly perform such fundamentalization, one needs a well-calibrated model and a knowledge of Cepheid's physical properties \citep[see, e.g.,][]{taormina2018puls, Gallenne_2018_V1334Cyg_Cep}, which are rarely available and, in most cases, are obtained from other theoretical (e.g., evolutionary) models. An interesting possibility would be to use the theoretically predicted periods for mixed-mode Cepheids in the OR regions to obtain similar fundamentalization relations as derived in this paper. Comparison with empirical relations could then be used to calibrate the models.

For RR Lyrae variables, metallicity depends significantly on the pulsation period, which intertwines with a dependence of period ratio on metallicity \citep{Marconi_2015_rrlyr_metal_dep, Chen_2023_rrly_per_perrat_metal}. This is not the case for Cepheids, where only a slight trend for a relation between [Fe/H] and period was detected with a low significance by \citet{daSilva_2022_MW_ceps_metal_scale_perdep}. There is, however, a hint in the data they presented that this trend may grow stronger for short periods. Unfortunately, it cannot be traced there because of scarce metallicity measurements for the faint, short-period Cepheids ($\log P_F < 0.5$, $\log P_{1O} < 0.35$). Our results, which show a different variation of the period ratios along the Cepheid period for each galaxy, can be considered another hint for metallicity trends and a more complex metallicity dependence. We note here that all our 1O/2O-mode Cepheids have $\log P_{1O} < 0.35$ (see Fig.~\ref{fig:fund2}). Additionally, the sensitivity to metallicity may depend on other Cepheid's physical properties and, as a result, on the period. A large spread of Cepheid masses and radii may thus be the cause of a change in the value and the sign of the metallicity dependence between the short and long-period Cepheids. A detailed theoretical study could help explain this phenomenon.

There is no direct metallicity determination for Cepheids in M31 and M33 galaxies, but indirect estimates exist. For example, \citet{WagnerK_2015_M31_cep_PLR_met} calculated [O/H] for Cepheids from their location in the disc using a relation of \citet{Zaritsky_1994_HII_spiral_galaxies} and obtained values that are close to solar.
For M33 \citet{Beaulieu_2006_M33_beat_cepheids}, estimated the metallicities of beat Cepheids using metallicity gradients and a comparison with theoretical models and obtained values of Z from about 0.006 to 0.013, which are consistent with the position of these Cepheids in Fig.~\ref{fig:fundcomp}.


\acknowledgments
The research leading to these results received funding from the Polish National Science Center grant SONATA BIS 2020/38/E/ST9/00486. This research used NASA's Astrophysics Data System Service.



\bibliography{fundper_letter}{}
\bibliographystyle{aasjournal}



\end{document}